\documentclass[twocolumn,showpacs,preprintnumbers,amsmath,amssymb]{revtex4}
\usepackage{graphicx}% Include figure files
\usepackage{dcolumn}% Align table columns on decimal point
\usepackage{bm}% bold math
\begin{document}

\preprint{CUPhysics/5/2008}
\title{Realistic searches on stretched exponential networks}

\author{Parongama Sen}
\affiliation{
Department of Physics, University of Calcutta,92 Acharya Prafulla Chandra Road,
Calcutta 700009, India.\\
}
%\email{psphy@caluniv.ac.in}

%\documentclass[twocolumn,showpacs,preprintnumbers,amsmath,amssymb]{revtex4}
%\documentclass[pre,showpacs,preprintnumbers,amsmath,amssymb]{revtex4}
%\mark{{Gallant king...}{X Y zzz and A B zzzz}}
%\email{psphy@caluniv.ac.in}

%\keywords{
%Small world effect, dynamic paths, social distances.}
%\pacs{89.75.Hc, 89.70.+c, 89.75.Fb}
%
%
%                                                                                                                             
%                                                                                                                             
%
%\abstract{
\begin{abstract}
We consider navigation or search schemes on networks which have  a degree distribution of the form 
$P(k) \propto \exp(-k^\gamma)$. In addition,  the linking probability 
is taken to be 
dependent on social distances and is governed by a parameter $\lambda$.
The searches  are realistic in the
sense that not all search chains can be completed.
An estimate of $\mu=\rho/s_d$, where $\rho$ is the success rate and 
$s_d$ the dynamic path length, shows that for a network of $N$ nodes,
  $\mu \propto N^{-\delta}$ in general.
Dynamic small world effect, i.e., $\delta \simeq 0$ is shown to exist in a restricted region of the $\lambda-\gamma$ plane.
\end{abstract}

                                                                                                                             
\maketitle
                                                                                                                            
\medskip

%\section{Introduction}

The subject of  networks has emerged as a multidisciplinary
field  in which 
 there has been tremendous activity in recent times \cite{watts,bareview}.
 The interest in networks has grown due to the fact that although  networks 
can be of diverse nature, 
there are some striking universal  properties in their
underlying structure.   
The most important property that appeared to be commonly
occurring in
networks is the small world property. This means  that if
any two nodes in the network is separated by an average number of $s$
steps,
 then $s \propto \ln(N)$, where $N$ is the total number of
nodes in the network.
In some networks, even slower variation (i.e., sub-logarithmic scaling) has been observed \cite{newman_sub}.

The first indication that networks have small world 
behaviour emerged from an experimental study by
Milgram et al \cite{milgram}, in which it was shown that 
any two persons (in the USA) can be connected by an
average number of six steps.
Following the tremendous interest in the study
of networks,  new experiments 
have been done to 
verify this property in real social networks \cite{killworth,dodds}.
Some studies which involve  simulations on real networks \cite{adamic_search,hong,geog}  
have been made also.
Parallely, the question of 
navigation on small world networks has been addressed theoretically
in many model networks \cite{klein,adamic1,kim,zhu,moura,watts-search,carmi,thada,clauset,sen1,sen2,leetal}. 

It must be noted that  it is not necessary that
a navigation or searching on a small world 
network would  show the small world property, i.e., the dynamic 
paths  $s_d$ may not scale as  $\ln(N)$.
This is because searching is done using local 
information only while the average shortest distances are calculated
using the global knowledge of the network. 
This was explicitly shown by Kleinberg \cite{klein} in a theoretical study
where nodes were placed on a two dimensional Euclidean space. 
Each node here has connections to its  nearest neighbours  as well as to
neighbours at a distance $l$ with probability  $ P(l) \propto l^{-\alpha}$.
Although the network is globally a small world for a range of values of 
$\alpha$, 
navigation on such networks using greedy algorithm showed a
small world behaviour only at $\alpha = 2$. 

Conventionally a network is said to be searchable if there exist short
dynamic paths scaling as $\ln(N)$. However, search chains have been shown 
to terminate unsuccessfully in many real
experiments. In theoretical studies, this has been considered
recently and it has been shown that with the possibility of failures, 
the scaling 
of the path lengths alone is not always a good measure of searchability \cite{sen1}.
The ratio of the success rate to the path lengths, on the other 
hand, gives a reliable measure. Based on this measure, the searchability on scale
free networks and Euclidean networks have been analysed recently \cite{sen1,sen2}.
The searchability depends on both the network structure and to
a large extent on the searching strategy.

In the present work, we have considered a network with a degree distribution
which is not scale-free but given by $P(k) \propto \exp({-k^\gamma})$, i. e., 
a stretched exponential behaviour.
Here,
we have a parameter $\gamma$ associated with the degree distribution;
a large $\gamma$ value indicates a very fast decay of $P(k)$ such that  the 
probability of 
 having high degree  nodes is very small. One the other hand, if $\gamma$ is small, such a probability may not be negligible even if the degree distribution
is not scale-free.
Moreover, we consider a characteristic feature attached to each node. 
The linking probability is taken to be dependent on 
the social distance or the difference of this characteristic feature between
nodes. 
Thus, this is an  attempt to  construct a simple model of social network.
A similar study was made in \cite{sen1} where a scale-free degree
distribution was used. However, in most social networks,
one has a degree distribution with a faster than algebraic decay and
therefore we have considered a stretched exponential degree distribution here. 

The network is generated by a method described in detail in
\cite{sen1}. In brief, we first assign the degrees to the nodes
according to the degree distribution. The characteristic feature called the
similarity factor is  
measured by a variable $\xi$ varying from 0 to 1 and is assigned to
each  node randomly. The edges are then introduced between pairs of nodes 
$i$ and $j$ with the probability

\begin{equation}
{\cal{P}}_{i,j} \propto |\xi_i -\xi_j|^{-\lambda},
\label{eq1}
\end{equation}
with $\lambda > 0$.

Thus we will have a  network 
in which similar nodes will try to link up for $\lambda > 0$.
A large value of $\lambda$ indicates that nodes which are very similar
tend to link up while for $\lambda=0$, there are no correlations
among nodes.
 The minimum and maximum allowed degrees are  two 
and  $N^{1/2}$ respectively.

We use an algorithm in which each  node  knows the similarity factor of its
immediate neighbours as well as that  of  the target node.
 Nodes send messages to
a neighbour most similar to the target node. Each node can receive the 
message only once. In case there is no node to which the 
message can be forwarded, the search will terminate.
Conducting the searches between arbitrary source-target pairs, we estimate
the success rates $\rho$ and the average dynamic paths $s_d$. The ratio
$\mu = \rho/s_d$ is then studied as a function of the network
size $N$.

We find that as in \cite{sen1,sen2}, $\mu$ has a power law decay with $N$
with an exponent $\delta$ 
for all $\gamma$ and $\lambda$ in general except for
very large values of $\lambda$ where there seems to be a correction to
the power law scaling. $\delta$ lies between zero and one, as found in the
previous studies. 
A smaller value of $\delta$
indicates a better searchability and in particular $\delta \simeq 0$ 
would correspond to a dynamic small world (DSW) effect \cite{sen1}.
For each $\gamma$, we  estimate $\delta$ 
as a function of  $\lambda$. 
We find that for small $\gamma$, there is indeed a range of values  of $\lambda$
where $\delta$ is very small, close to zero. 
On the other hand,
for $\gamma > \gamma_c$, where $\gamma_c$ is close to 0.4, 
there is no such region. 
%Therefore,
%one can conceive of a $\lambda-\gamma$ plane where 
%there is a bounded region of dynamic small world effect.
In the present study,
if $\delta < 0.1$ (which means the decay could as well be logarithmic), 
we assume that there is a dynamic small world effect. 

\begin{figure}
%\epsfxsize=7cm
%\centerline{\epsfbox{weight_gam4muvsN.eps}}
\includegraphics[clip,width= 7cm, angle=270]{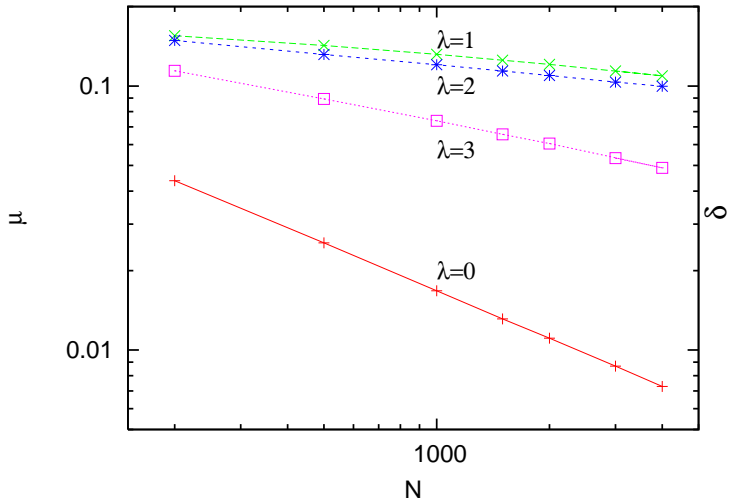}
\caption{The typical variations of the ratio $\mu$ of the success rate to the 
dynamic path lengths as a function of $N$ are shown for different
values of $\lambda$ for $\gamma=0.4$. The slopes give the 
estimate of $\delta$.}
\end{figure}

Fig.1 shows typical variations  of $\mu$ versus $N$ for 
$\gamma=0.4$ for various values of $\lambda$. The slope of
these curves gives the exponent $\delta$. 
In Fig. 2, $\delta$ against $\lambda$ values are shown for various $\gamma$
where we observe 
that for low values of $\gamma$, there
is a finite region where $\delta$ is less than 0.1. 
This region shrinks as $\gamma$ is increased.
However, even when there is no dynamic small world effect, we find that
$\delta$ reaches a minimum at $\lambda=\lambda_{min}$ where $\lambda_{min} \simeq 1.25$ independent of  $\gamma$. Hence for a stretched exponential
network in general, where the linking probability is dependent 
on the social distances parametrically as in (\ref{eq1}), the network is most searchable for a fixed value of the parameter.
 
 We have schematically shown the 
DSW region in the $\lambda-\gamma$ plane in Fig. 3. We would like to mention here  that
for the social network considered earlier, with a  power law  degree distribution ($P(k) \propto k^{-2})$, 
DSW had been obtained for a finite  range of value of $\lambda$ as well \cite{sen1}.

Thus we conclude that for realistic searches on social networks,
where the degree distribution is given by a stretched exponential
function, one can achieve dynamic small world effect as long as the
degree distribution does not decay very fast and the network is
neither too correlated nor too random as far as the similarity of nodes
is concerned. In general, we find that while the quality of the 
searchability is 
determined by $\lambda$,  $\gamma$ makes it
quantitatively different, e.g.,   at smaller
values of $\gamma$ where there is a larger number of highly
connected nodes, the searchability increases.

We also find that in the DSW region, 
the success rate is almost a constant as a function of 
$N$, and the  dynamic path lengths  increase very slowly indicating
a logarithmic increase. The absolute value of path lengths is also ``small'',
e.g., for a network of size 4000, the dynamic path length 
 is about 5 in the DSW region. 
 This could explain the observed results of searching in real
networks.  Of course, in a real social search, a 
 much more complicated dynamics is involved,
with nodes having more than one characteristic feature and each node
using its own searching strategy.

\begin{figure}
%\epsfxsize=7cm
%\centerline{\epsfbox{delta_vs_lam.eps}}
\includegraphics[clip,width= 7cm, angle=270]{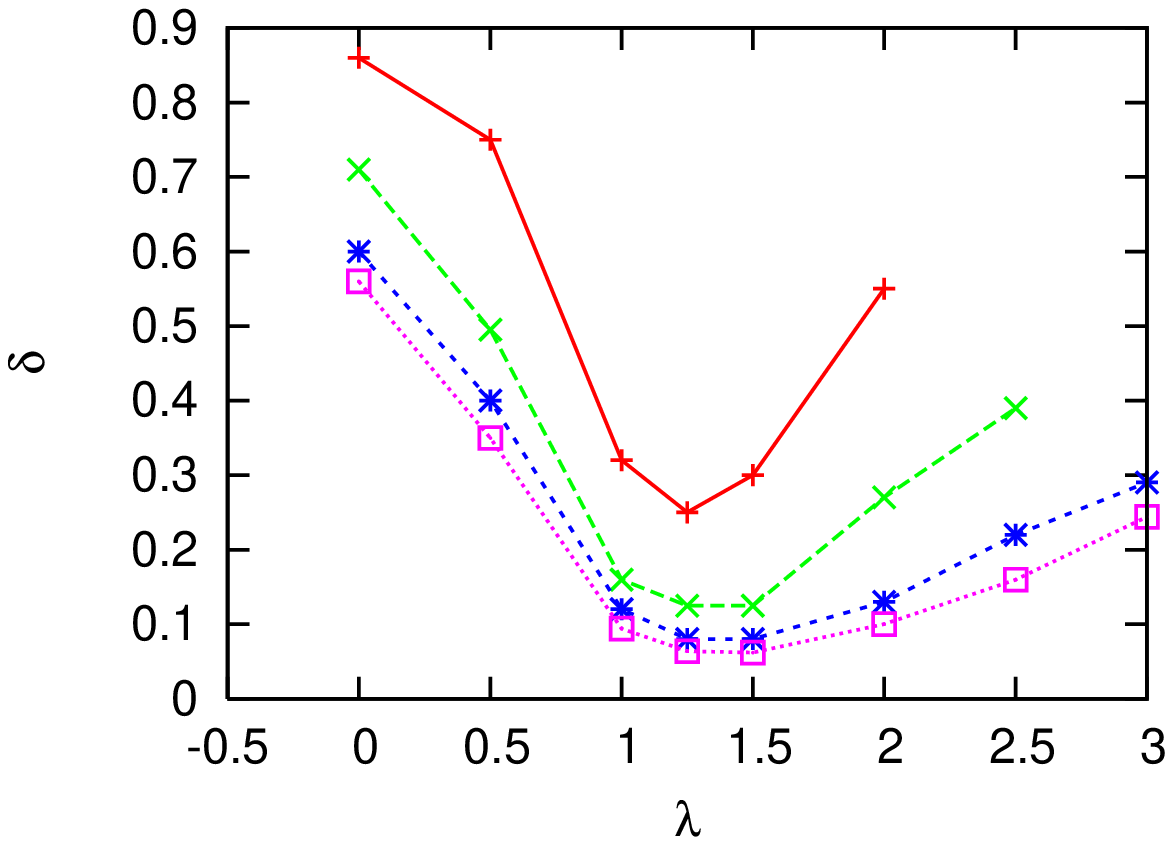}
\caption{The values of $\delta$ against $\lambda$  
 show that for low  values of $\gamma$, a dynamic small world effect 
 can be achieved
for a finite interval of $\lambda$ values. The curves are shown for $\gamma=0.6
, \gamma=0.5, \gamma=0.4$ and $\gamma=0.3$, from top to bottom.}
\end{figure}
\begin{figure}
%\epsfxsize=7cm
%\centerline{\epsfbox{dsw.eps}}
\includegraphics[clip,width= 7cm]{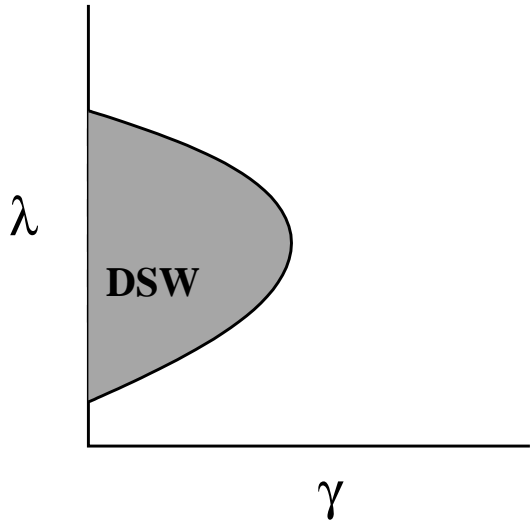}
\caption{Schematic picture of  the $\lambda-\gamma$ plane
showing the region with the dynamic small world effect.}
\end{figure}

Acknowledgement: Financial support from CSIR grant no. 3(1029)/05-EMR-II
  is acknowledged.


\begin{thebibliography}{99}

\vskip -4cm

\bibitem{watts}
 D. J. Watts and S. H. Strogatz, Nature {\bf 393}, 440 (1998);
  D. J. Watts, {\it Small Worlds}, Princeton Univ. Press,
   Princeton (1999).
\bibitem{bareview}   R. Albert and A.-L. Barab\'asi,
Rev. Mod. Phys. {\bf 74}, 47 (2002).
\bibitem{newman_sub} M. E. J. Newman,
SIAM Rev. {\bf 45},  167 (2003).

\bibitem{milgram} 
S. Milgram, 
      {Psychology Today}  {\bf 1}, 60  ({1967});
 J. Travers and S.   Milgram, {Sociometry} {\bf 32}, 425 ({1969}).
\bibitem{killworth}P. D. Killworth and H. R. Bernard,   Social Networks
 {\bf 1}159 (1978).
\bibitem{dodds}
 {P. S. Dodds,  R. Muhamad and  D. J. Watts},
{Science}      {\bf     301}, 827   (2003).
\bibitem{adamic_search}
{L. A. Adamic and E.  Adar}, {Social Networks} {\bf 27}, 187  (2005).
\bibitem{hong} I. Clarke, S. G. Miller, T. W. Hong, O. Sandberg and 
B. Wiley, IEEE Internet Computing {\bf 6}, 40 (2002).
\bibitem{geog} D. Liben\_Nowell, J. Novak, R. Kumar, P.Raghavan and A. Tomkins,
PNAS {\bf 102}, 11623 (2005).
\bibitem{klein}   J. Kleinberg, Nature {\bf 406}, 845 (2000).
\bibitem{adamic1}
 {L. A. Adamic,  R. M. Lukose,   A. R. Puniyani and  B. A. Huberman},
 {Phys. Rev. E} {\bf 64}, 041235 (2001).
\bibitem{kim} B. J. Kim, C. N. Yoon, S. K. Han and H. Jeong, Phys. Rev. E
{\bf 65}, 027103 (2002). 
\bibitem{zhu} {H. Zhu and  Z-X. Huan},  {Phys. Rev. E} {\bf 70} 036117 (2004).
\bibitem{moura} {A. P. S. de Moura, A. E. Motter and C. Grebogi},
{Phys. Rev. E} {\bf {68}} 036106 (2003). 
\bibitem{watts-search}
{D. J. Watts,  P. S. Dodds and  M. E. J. Newman}, {Science}  {\bf 296}, 1302 
(2002).
\bibitem{carmi} 
 S. Carmi, R. Cohen and D. Dolev,
 Europhys. Lett. {\bf 74},  1102 (2006).
\bibitem{thada} 
 H. P. Thadakamalla, R. Albert and  S. R. T. Kumara,
Phys. Rev. E {\bf 72}, 066128 (2005).
\bibitem{clauset} A. Clauset and C. Moore, preprint arxiv:cond-mat/0309415.
%   \bibitem{BA} A.-L. Barab\'asi and R. Albert, Science {\bf 286}, 509 (1999).
%\bibitem{latora} V. Latora and M. Marchiori, Phys. Rev. Lett. {\bf 87}
 198701 (2001).
\bibitem{sen1}
 P. Sen, J. Stat. Mech. P04007 (2007).
\bibitem{sen2} K. Basu Hajra and P. Sen,
 J. Stat. Mech.  P06015  (2007).
\bibitem{leetal} S. H. Lee, P-J. Kim, Y-Y. Ahn and H. Jeong, arXiv:0710.3268v1.
%\bibitem{holme} 
%P. Holme, Phys. Rev. E {\bf 71}  046119 (2005).
%\bibitem{lazaros} L. K. Gallos and P. Argyrakis, Physica A {\bf 330} 117 (2003).
%\bibitem{newman_asso} 
%M. E. J. Newman, {Phys. Rev. Lett.}
%{\bf 89},  208701 (2002). 
%\bibitem{amaral}L. A. N Amaral, A. Scala, M. Barthelemy and H. E. Stanley,
%Proc. Nat. Acad. Sc. {\bf 97} 11149 (2000).


\end{thebibliography}
\end{document}